\documentclass{arxiv}
\usepackage{graphicx}
\usepackage{eurosym}
\usepackage{hyperref}
\usepackage{tablefootnote}

\newcommand{\keuro}{\mbox{k\euro}}
\newcommand{\meuro}{\mbox{M\euro}}
\newcommand{\co}{\mbox{CO$_2$}}
\newcommand{\coe}{\mbox{CO$_2$e}}
\newcommand{\tco}{\mbox{tCO$_2$}}
\newcommand{\kgcoe}{\mbox{kgCO$_2$e}}
\newcommand{\tcoe}{\mbox{tCO$_2$e}}
\newcommand{\ktcoe}{\mbox{ktCO$_2$e}}
\newcommand{\Mtco}{\mbox{MtCO$_2$}}
\newcommand{\Mtcoe}{\mbox{MtCO$_2$e}}
\newcommand{\tcoeyr}{\mbox{tCO$_2$e yr$^{-1}$}}
\newcommand{\ktcoeyr}{\mbox{ktCO$_2$e yr$^{-1}$}}

\newcommand{\moneyeft}{\mbox{tCO$_2$e \meuro$^{-1}$}}
\newcommand{\masseft}{\mbox{tCO$_2$e kg$^{-1}$}}

\newcommand{\apjl}{Astrophys. J. Lett.}   
\newcommand{\aap}{Astron. Astrophys.}   
\newcommand{\aph}{Astropart. Phys.} 
\newcommand{\baas}{Bull. Am. Astron. Soc.}   
\newcommand{\eastro}{Exp. Astron.} 
\newcommand{\nat}{Nature} 
\newcommand{\nastro}{Nat. Astron.} 
\newcommand{\pnas}{Proc. Natl Acad. Sci. USA}   
\newcommand{\procspie}{Proc. SPIE}   
\title{Estimate of the carbon footprint of astronomical research infrastructures}

\author{J\"urgen Kn\"odlseder$^{1}$, Sylvie Brau-Nogu\'e$^{1}$, Mickael Coriat$^{1}$,
            Philippe Garnier$^{1}$, Annie Hughes$^{1}$, Pierrick Martin$^{1}$ \&
            Luigi Tibaldo$^{1}$}

\begin{document}
\maketitle
\spacing{1.5}

\begin{affiliations}
\item Institut de Recherche en Astrophysique et Plan\'etologie, Universit\'e de Toulouse, CNRS, CNES, UPS,
9 avenue Colonel Roche, 31028 Toulouse, Cedex 4, France
\end{affiliations}

\abstract{
The carbon footprint of astronomical research is an increasingly topical issue with first estimates of
research institute and national community footprints having recently been published. 
As these assessments have typically excluded the contribution of astronomical research infrastructures, 
we complement these studies by providing an estimate of the contribution of astronomical space missions 
and ground-based observatories using greenhouse gas emission factors that relates cost and payload 
mass to carbon footprint. 
We find that worldwide active astronomical research infrastructures currently have a carbon footprint of
20.3$\pm$3.3 \Mtco\ equivalent (\coe) and an annual emission of 1,169$\pm$249 \ktcoeyr\ corresponding 
to a footprint of 36.6$\pm$14.0 \tcoe\ per year per astronomer. 
Compared with contributions from other aspects of astronomy research activity, our results suggest that research 
infrastructures make the single largest contribution to the carbon footprint of an astronomer. 
We discuss the limitations and uncertainties of our method and explore measures that can bring greenhouse 
gas emissions from astronomical research infrastructures towards a sustainable level.
}
\newline
\normalfont


The Sixth Assessment Report of the Intergovernmental Panel on Climate Change (IPCC) Working Group I
could not be more explicit: ``It is unequivocal that human influence has warmed the atmosphere, 
ocean and land [...] Global warming of 1.5$^\circ$C and 2$^\circ$C will be exceeded during the 21st 
century unless deep reductions in \co\ and other greenhouse gas emissions occur in the coming 
decades.''\cite{ipcc2021}.
As stated by Ant\'{o}nio Guterres, United Nations Secretary-General, the IPCC Sixth Assessment Report is a 
``code red for humanity''\cite{guterres2021}.

Taking up this code-red alert, there is growing recognition in the astrophysics community that it 
must assume its share of the global effort to reduce greenhouse gases (GHGs)\cite{marshall2009, 
matzner2019, williams2019, stevens2020}.
Much recent attention has focused on the reduction of academic flying\cite{lequere2015, 
rosen2017, hamant2019, burtscher2020, barret2020}
and, to a lesser extent, on the use of supercomputers\cite{stevens2020, jahnke2020}.
Quantifying the GHG emissions due to the construction and operation of space observatories, planetary 
probes and ground-based observatories has so far attracted less attention. 
With the decades-long lifetime of research infrastructures, decisions that are made now will lock in 
GHG emissions of the astrophysics community for the next decades, potentially compromising the 
goal to reach net-zero emissions by the middle of this century. 
Assessments of the environmental footprint of existing and future astronomical facilities 
are therefore urgently needed.

To address this gap, we have developed a method that provides a first-order estimate of the carbon 
footprint of astronomical research infrastructures (see `Carbon footprint estimation' in Methods). 
We estimate the carbon footprint of each facility using the standard method of multiplying activity 
data with emission factors. 
Detailed activity data (for example, MWh of electricity consumed for construction of a satellite or 
tonnes of concrete poured for the foundation of a telescope) are generally not publicly available for 
astronomical research infrastructures, so we use aggregated activity data based on cost and mass 
for our analysis. 
Specifically, we use the full mission cost and payload launch mass for space missions and the 
construction and operating costs for ground-based observatories (Supplementary Tables 1 and 2).

Using cost data to estimate a project's carbon footprint is referred to as economic input--output (EIO) 
analysis. 
This approach is known to have large uncertainties due to the aggregation of activities, products and 
monetary flows that may vary considerably from one facility or field of activity to another. 
An alternative life-cycle assessment (LCA) methodology is recommended by key space industry 
actors (for example, ref.~13) as the optimal method to assess and reduce the carbon footprint 
of space missions, but it is difficult to implement in practice (especially for comparative or discipline-wide 
assessments) due to the confidential nature of the required input activity data\cite{maury2020}.
At present, an EIO analysis is thus the only feasible way to assess the combined carbon footprint of the 
world's space- and ground-based astronomical research infrastructures. 
We adopt throughout this study an uncertainty of 80\% for the carbon footprint estimate of individual 
facilities, as recommended by the French Agency for Ecological Transition (ADEME) for an EIO 
analysis\cite{basecarbone}.

Detailed carbon footprint assessments do exist for a handful of facilities, and we use these assessments 
to derive dedicated emission factors for our analysis. 
We emphasize that our results in this paper are order-of-magnitude estimates that may differ by a factor 
of a few from the true carbon footprint of any given facility. 
We propagate uncertainties throughout our analysis and the aggregated results and their associated 
uncertainties are robust. 
We note that our total and per-astronomer estimates are likely to be conservative, since we simply excluded
any activity data that we could not locate, and our estimate for the number of astronomers in the world is 
probably an upper limit (see `The number of astronomers in the world' in Methods). 
The average carbon footprint of astronomical research infrastructures per astronomer is thus likely 
to be larger than our estimate, or towards the upper bound of our quoted uncertainty interval.

Our work was conducted in the context of the carbon footprint assessment of our institute, the 
Institut de Recherche en Astrophysique et Plan\'etologie (IRAP) for the year 2019 and hence we 
adopt 2019 as the reference year for our study. 
Specifically, all cost data have been corrected for inflation and are expressed in 2019 economic 
conditions. 
We based our work on a list of 46 space missions and 39 ground-based observatories from which 
data were used to produce peer-reviewed journal articles authored or co-authored by IRAP scientists 
in 2019. 
Extrapolating the results for this list to all active astronomical research infrastructures in the world 
yields an estimate for the worldwide carbon footprint of astronomical facilities.

\section*{Results}

To estimate emission factors for astronomical research infrastructures, we made a comprehensive 
search for published carbon footprint assessment reports. 
As detailed in `Emission factors' in Methods, these reports are currently very scarce. 
We found two case studies for life-cycle carbon footprints of space missions, which covered the entire 
mission including the launcher and a few years of operations\cite{wilson2019}. 
From these studies, we infer mean emission factors of 140 \tco\ equivalent (\coe) per million \euro\ (\meuro) of mission cost 
and 50 \masseft\ of payload launch mass. 
Emission factors of ground-based observatories were derived using existing carbon footprint 
assessments for the construction of two facilities and the operations of three facilities. 
We find a mean emission factor of 240 \moneyeft\ for construction and of 250 \moneyeft\ for operations. 
The emission factors are summarized in Table \ref{tab:emission-factors}. 
The lower monetary emission factor for space missions compared with ground-based observatories 
can be attributed to the low production rates, long development cycles and specialized materials and 
processes of the space sector\cite{geerken2018}.

\begin{table}
\caption{Adopted emission factors.
\label{tab:emission-factors}}
\centering
\begin{tabular}{l c}
\hline
Activity & Emission factor \\
\hline
Space missions (per payload launch mass) & 50 \masseft\ \\
Space missions (per mission full cost) & 140 \moneyeft\ \\
Ground-based observatory construction & 240 \moneyeft\ \\
Ground-based operations & 250 \moneyeft\ \\
\hline
\end{tabular}
\end{table}

Table \ref{tab:spacefootprint} summarizes order-of-magnitude estimates for the carbon life-cycle footprints of 
space missions based on payload launch mass and mission cost 
(see `Carbon footprint estimation' and `Emission factors' in Methods and Supplementary Tables 1 and 2). 
The cost-based estimates are on average about 20\% larger than the mass-based estimates, 
probably because mission complexity and mission extensions are not taken into account by the latter. 
Examples include the Hubble Space Telescope (HST), which had five Space Shuttle 
servicing missions that are included in the cost-based estimate but not in the mass-based 
estimate, and the Mars rover Curiosity, for which the mission complexity is not properly 
reflected in the mass-based estimate.

\begin{table}
\tiny
\caption{Order of magnitude estimates of the carbon footprints of some selected space
missions, ordered by decreasing mass-based footprint.
For cells that are blank no cost estimates could be found.
\label{tab:spacefootprint}}
\centering
\setlength\tabcolsep{3pt}
\vspace{6pt}
\begin{tabular}{l r r r r r r r r r r r r r r}
\hline
Mission & Years & Papers & Authors && \multicolumn{4}{c}{Mass-based} && \multicolumn{4}{c}{Cost-based} \\
\cline{2-4} \cline{6-9} \cline{11-14}
& \multicolumn{3}{c}{total since launch} && Footprint & Annual & \multicolumn{2}{c}{Carbon intensity} && Footprint & Annual & \multicolumn{2}{c}{Carbon intensity} \\
& & & && (\tcoe) & (\tcoeyr) & (\tcoe\ per & (\tcoe\ per && (\tcoe) & (\tcoeyr) & (\tcoe\ per & (\tcoe\ per \\
& & & &&  & & paper) & author) && & & paper) & author) \\
\hline
HST               & 30 & 52,497 & 42,315 && 555,500 & 18,517 &  11 & 13   && 1,125,197 &37,507 & 21 & 27 \\
Chandra        & 21 & 17,714 & 23,942 && 293,000 & 13,952 &  17 & 12   && 575,955 & 27,426 & 33 & 24 \\
Cassini          & 22 &   4,691 &   9,328 && 291,000 & 13,227 &  62 & 31   && 392,902 & 17,859 & 84 & 42 \\
Cluster           & 20 &   2,433 &  2,959 && 240,000 & 12,000 &  99 & 81    && 132,207 & 6,610 & 54 & 45 \\
Fermi             & 12 &   8,619 & 19,675 && 215,150 & 17,929 &  25 & 11   && 120,881 & 10,073 & 14 & 6 \\
INTEGRAL    & 18 &   2,808 & 10,640 && 200,000 & 11,111 &  71 & 19    && 58,720 & 3,262 & 21 & 6 \\
Curiosity        &  7 &   1,360 &    4,393 && 194,650 & 19,465 & 143 & 44  && 362,595 & 36,259 & 267 & 83 \\
XMM              & 21 & 18,859 & 23,773 && 190,000 & 9,048 &  10 & 8       && 155,845 & 7,421 & 8 & 7 \\
Juno              &   8 &       521 &   1,832 && 181,250 & 18,125 & 348 & 99   && 151,547 & 15,155 & 291 & 83 \\ 
Herschel        & 11  &   5,046 & 11,092 && 170,000 & 15,455 &  34 & 15   && 161,238 & 14,658 & 32 & 15 \\
RXTE             & 24 &   7,473 & 11,601 && 160,000 & 6,667 &  21 & 14     &&  50,438 & 2,102 & 7 & 4 \\
SDO               & 10 &   4,189 &   4,946 && 155,000 & 15,500 &  37 & 31   &&  121,164 & 12,116 & 29 & 24 \\
Rosetta          & 16 &   1,665 &   4,337 && 145,000 & 9,063 &  87 & 33      && 239,316 & 14,957 & 144 & 55 \\
Galileo            & 30 &   2,432 &  4,594 && 128,000 & 4,267 &  53 & 28      && 178,503 & 5,950 & 73 & 39 \\
MAVEN           &   6 &       672 &  2,023 && 122,700 & 12,270 & 183 & 61   && 89,270 & 8,927 & 133 & 44 \\
ROSAT           & 30 & 19,765 & 23,154 && 121,050 & 4,035 &    6 & 5       &&  88,844 & 2,961 & 4 & 4 \\
MRO               & 14 &   1,927 &  4,261 && 109,000 & 7,786 &   57 & 26     &&  129,850 & 9,275 & 67 & 30 \\
Gaia                &   7 &   2,550 & 10,565 && 101,700 & 10,170 &  40 & 10   && 145,114 & 14,511 & 57 & 14 \\
Planck             & 11 &  5,515  & 13,388 && 95,000 & 8,636 &   17 & 7        && 108,486 & 9,862 & 20 & 8 \\
SoHO              & 25 & 12,218  & 12,955 && 92,500 & 3,700 &     8 & 7       && 205,617 & 8,225 & 17 & 16 \\
Suzaku            & 15 &   3,869  &   9,525 && 85,300 & 5,687 &   22 & 9       && & & & \\
AstroSat          &  5 &        313  &   5,406 && 75,750 & 7,575 & 242 & 14     && 3,751 & 375 & 12 & 1 \\
MMS                & 5 &         769 &   1,623 && 68,000 & 6,800 &   88 & 42      && 147,501 & 14,750 & 192 & 91 \\
Venus Express & 15 &  1,221 &   3,394 && 63,500 & 4,233 &   52 & 19      &&  41,945 & 2,796 & 34 & 12 \\
Wind                & 26 &  3,877 &    8,254 && 62,500 & 2,404 &   16 & 8        && & & & \\
STEREO          & 14 &  3,731 &    6,768 && 61,900 & 4,421 &   17 & 9       && 86,021 & 6,144 & 23 & 13 \\
Mars Express   & 17 &  2,969 &   6,118 && 61,150 & 3,597 &    21 & 10      && 52,332 & 3,078 & 18 & 9 \\
Dawn                & 12 &      791 &   2,175 && 60,885 & 5,074 &   77 & 28       && 61,409 & 5,117 & 78 & 28 \\
Hipparcos         & 31 &  4,743 &   8,373 && 57,000 & 1,839 &   12 & 7         && 130,664 & 4,215 & 28 & 16 \\
Kepler               & 11 &  4,306 &   9,606 && 52,620 & 4,784 &   12 & 5         && 89,037 & 8,094 & 21 & 9 \\
Geotail             & 28 &  3,288 &   3,996 && 50,450 & 1,802 &   15 & 13       && & & & \\
Akari                 & 14 &  2,037 &   6,993 && 47,600 & 3,400 &   23 & 7         && 14,878 & 1,063 & 7 & 2 \\
Spitzer              & 17 &  9,050 & 15,940 && 47,500 & 2,794 &     5 & 3         && 166,333 & 9,784 & 18 & 10 \\
Swift                 & 16 &  7,397 & 17,307 && 42,150 & 2,634 &     6 & 2         && 39,030 & 2,439 & 5 & 2 \\
ACE                  & 23 &  4,147 &   7,560 && 37,600 & 1,635 &     9 & 5         && & & & \\
InSight              &   1 &        58 &       447 && 34,700 & 3,470 & 598 & 78       && 99,922 & 9,992 & 1723 & 224 \\
PSP                  &   2 &      287 &   1,075 && 34,250 & 3,425 & 119 & 32       && 183,456 & 18,346 & 639 & 171 \\ 
WISE                & 11 &  6,990 & 18,877 && 33,050 & 3,005 &     5 & 2         && 46,855 & 4,260 & 7 & 2 \\
TIMED              & 18 &  2,205 &   3,593 && 33,000 & 1,833 &    15 & 9        && 36,196 & 2,011 & 16 & 10 \\
Double Star      & 16 &      166 &       540 && 28,000 & 1,750 &  169 & 52      && & & & \\
IMP-8               & 47 &  2,485 &    3,835 && 20,500 &    436 &       8 & 5       && & & & \\
NICER             &   3 &      338 &    2,657 && 18,600 & 1,860 &    55 & 7        && 7,374 & 737 & 22 & 3 \\
NuSTAR          &   8 &  2,227 &    9,559 && 18,000 & 1,800 &      8 & 2        && 21,799 & 2,180 & 10 & 2 \\
TESS               &   2 &     978 &    4,557 && 16,250 & 1,625  &    17 & 4        && 38,478 & 3,848 & 39 & 8 \\
GALEX            & 17 &  5,452 &  13,790 && 14,000 &    824  &      3 & 1        && 16,780 & 987 & 3 & 1 \\
DEMETER       & 16 &     422 &     1,014 && 6,500  &     406 &     15 & 6        && 2,907 & 182 & 7 & 3 \\
\hline
\end{tabular}
\end{table}

For some missions, the mass-based estimates are larger. 
In some cases, this can be explained by lower mission complexity, but it may also be due to underestimates 
of the true costs. 
For example, the International Gamma-Ray Astrophysics Laboratory (INTEGRAL) makes use of 
the same satellite bus as the X-ray Multi-Mirror mission (XMM-Newton), which led to important 
cost savings. 
On the other hand, the quoted cost estimate for INTEGRAL only covers the mission cost to the 
European Space Agency (ESA), 
excluding the payload cost and the launcher cost, the latter having been provided by Russia in 
exchange for observing time. 
The Cluster mission consists of four identical spacecrafts and shares the same launcher for all 
four satellites, which also results in a relatively low cost-to-mass ratio. 
We also note that cost estimates for the Astronomy Satellite (AstroSat) and Akari appear to be substantially underestimated.

We consider that using both mass-based and cost-based estimates gives a good indication of the 
typical uncertainty that is inherent to our approach. 
ESA advises against using EIO analyses due to their large 
inherent uncertainties\cite{esahandbook}, yet our analysis suggests that as long as the uncertainties 
are properly considered, an EIO analysis for space missions yields results that are comparable with an 
analysis based on payload mass. 
Summing the carbon footprint estimates for the 40 space missions that have both mass-based and 
cost-based estimates yields 4.6$\pm$0.8 \Mtcoe\ for the mass-based and 5.9$\pm$1.2 \Mtcoe\ 
for the cost-based estimates, where errors reflect the adopted uncertainty of 80\% in 
the carbon footprint of each individual space mission (uncertainties for individual 
facilities are added in quadrature through the paper by taking the square root of the sum of 
uncertainty squares). 
Summing the mass-based estimates for all 46 space missions increases the total carbon footprint by 6\%
to 4.9$\pm$0.8 \Mtcoe. 
Assuming that the same increase would apply to the remaining 6 space missions for which we did not 
find any cost estimates would lead to a cost-based estimate of 6.2$\pm$1.3 \Mtcoe. 
Table \ref{tab:spacefootprint} also lists the annual footprints of the facilities, computed by dividing the 
life-cycle footprints by the lifetime of the mission, defined as the time since launch or ten years, whichever 
is longer (see `Annual footprint' in Methods). 
The total annual footprint of the missions in Table \ref{tab:spacefootprint} is 310$\pm$47 \ktcoeyr\ for the 
mass-based and 366$\pm$64 \ktcoeyr\ for the cost-based estimates.

\begin{table}
\tiny
\caption{Order of magnitude estimates of the carbon footprint of some selected ground-based
astronomical observatories or telescopes, ordered by decreasing footprint over the lifetime of the
infrastructure.
For cells that are blank no cost estimates could be found.
\label{tab:groundfootprint}}
\centering
\setlength\tabcolsep{3pt}
\vspace{6pt}
\begin{tabular}{l r r r r r r r r r r}
\hline
Observatory & Lifetime (yr) & Papers & Authors & \multicolumn{4}{c}{Footprint} & & \multicolumn{2}{c}{Carbon intensity} \\
\cline{5-8} \cline{10-11}
& & & & Construction & Operation & Lifetime & Annual & & & \\
& & & & (\tcoe) & (\tcoeyr) & (\tcoe) & (\tcoeyr) & & (\tcoe\ per & (\tcoe\ per \\
& & & &  &  &  &  & & paper) & author) \\
\hline
VLT (Paranal)                    &  21 & 17,235 & 26,442 & 332,280 & 9,875 & 539,655 & 25,698 && 31 & 20 \\
ALMA                                &   9  &   7,460 & 18,610 & 299,576 & 26,196 & 535,340 & 56,154 && 72 & 29 \\
SOFIA                               &   9   &      662 &   3,586 & 263,544 & 22,375 & 464,919 & 48,729 && 702 & 130 \\
AAT                                   &  46  &  4,297 & 10,848 & 29,728 & 3,824 & 205,610 & 4,470 && 48 & 19 \\
VLA                                   &  40 & 26,918 & 28,206 & 82,817 & 2,400 & 178,826 & 4,471 && 7 & 6 \\
VLBA                                 &  27 &   4,995 & 12,427 & 31,608 & 3,874 & 136,194 & 5,044 && 27 & 11 \\
IRAM                                 &  30 &   6,744 & 12,095 & 12,240 & 3,750 & 124,740 & 4,158 && 18 & 10 \\
Gemini-South                    &  20 &   1,735 &   9,949  & 32,280 & 3,250 & 97,280 & 4,864 && 56 & 10 \\
CFHT                                 &  41 &   8,400 & 16,228 & 20,414 & 1,575 & 84,989 & 2,073 && 10 & 5 \\
ESO 3.6m (La Silla)           &  43 &   3,774 &    8,515 & 23,815 & 1,298 & 79,608 & 1,851 && 21 & 9 \\
GBT                                  &   19 &    2,554 &   9,905 & 28,812 & 2,436 & 75,088 & 3,952 && 29 & 8 \\
LOFAR                              &    8 &    2,205 & 10,304 & 48,000 & 2,291 & 66,326 & 7,091 && 30 & 6 \\
JCMT                                 &  33 &   4,726 &   9,145 & 9,192 & 1,364 & 54,194 & 1,642 && 11 & 6 \\
ATCA                                 &  32 &    4,108 & 12,537 & 22,863 & 873 & 50,787 & 1,587 && 12 & 4 \\
H.E.S.S.                            &  17 &    4,577 &  12,889 & 11,848 & 2,193 & 49,126 & 2,890 && 11 & 4 \\
MeerKAT                           &    2 &        335 &   2,750 & 30,624 & 3,190 & 37,004 & 6,252 && 110 & 13 \\
GTC                                  &  11 &    1,059 &    6,445 & 29,880 & & 29,880 & 2,716 && 28 & 5 \\
NRO                                  &  38 &    1,776 &   3,739 & 12,233 & 378 & 26,609 & 700 && 15 & 7 \\
LMT                                   &   6 &        213 &    1,912 & 18,504 & 786 & 23,221 & 2,637 && 109 & 12 \\
MLSO                                &  55 &       385 &        932 &  & 306 & 16,817 & 306 && 44 & 18 \\
APEX                                 &  15 &   2,244 &    8,097 & 4,800 & 675 & 14,925 & 995 && 7 & 2 \\
SMA                                   &  17 &   1,585 &    5,312 & 14,354 & & 14,354 & 844 && 9 & 3 \\
EHT                                   &  11 &       606 &    2,079 & 12,580 & - & 12,580 & 1,144 && 21 & 6 \\
Noto Radio Observatory    &  32 &      108 &     1,490 &  & 378 & 12,096 & 378 && 112 & 8 \\
2m TBL                               &  40 &     435 &     1,392 & 1,435 & 250 & 11,435 & 286 && 26 & 8 \\
2.16m (Xinglong Station)   &  46 &      235 &         651 & 1,750 & 182 & 10,137 & 220 && 43 & 16 \\
1.93m OHP                        &  62 &      394 &     2,056 & 1,309 & 136 & 9,763 & 157 && 25 & 5 \\
KMTNet                              &    5 &      169 &     4,191 & 4,193 & 437 & 6,377 & 856 && 38 & 2 \\
THEMIS                              &  21 &      142 &        307 &   & 275 & 5,775 & 275 && 41 & 19 \\
2.4m LiJiang (YAO)            &  12 &      149 &         688 & 2,297 & 239 & 5,168 & 431 && 35 & 8 \\
2m HCT                              &  19 &      276 &     1,259 & 1,454 & 151 & 4,331 & 228 && 16 & 3 \\
1.5m Tillinghast (FLWO)     &  51 &      652 &    2,514 & 683 & 71 & 4,312 & 85 && 7 & 2 \\
1.5m (OAN-SPM)               &  50 &      253 &    1,258 & 683 & 71 & 4,241 & 85 && 17 & 3 \\
1.8m (BOAO)                      &  24 &     262 &        892 & 1,093 & 114 & 3,827 & 159 && 15 & 4 \\
1m (Pic-du-Midi)                  &  57 &       29 &        345 & 240 & 25 & 1,665 & 29 && 57 & 5 \\
1.3m Warsaw (OGLE)         &  23 & 4,210 &    9,470 & 472 & 49 & 1,604 & 70 && 0.4 & 0.2 \\
C2PU                                   &   6 &        31 &    1,982 & 480 & 50 & 780 & 98 && 25 & 0.4 \\
TAROT                                &  22 &       206 &   5,602 & 216 & 23 & 711 & 32 && 3 & 0.1 \\
1m NOWT                           &   7 &         17 &         118 & 240 & 25 & 415 & 49 && 24 & 4 \\
\hline
\end{tabular}
\end{table}

Order-of-magnitude estimates for the carbon life-cycle footprints of ground-based observatories are summarized 
in Table \ref{tab:groundfootprint}. 
After a few decades, the operations footprint dominates the life-cycle footprint. 
The life-cycle carbon footprint of the ground-based observatories listed in Table \ref{tab:groundfootprint} is 
estimated to be 3.0$\pm$0.8 \Mtcoe, with a total footprint for construction of 1.4$\pm$0.4 \Mtcoe\ (46\%)
and for operations of 1.6$\pm$0.4 \Mtcoe\ (54\%). 
The total annual carbon footprint amounts to 194$\pm$64 \ktcoeyr\ for the ground-based 
observatories in Table \ref{tab:groundfootprint}.

To put the results in perspective, we compute the carbon intensity of each infrastructure, defined 
as the lifetime footprint divided by either the number of peer-reviewed papers or the number of unique 
authors (see `Carbon intensity' in Methods). 
The meaning of these quantities is that they relate the total carbon footprint of a given infrastructure 
to the scientific productivity of the community and to the size of the community that makes use of it. 
Specifically, the latter provides a measure of how the infrastructure footprint is shared among the user 
community.

\begin{figure}
\centering
\includegraphics[width=15cm]{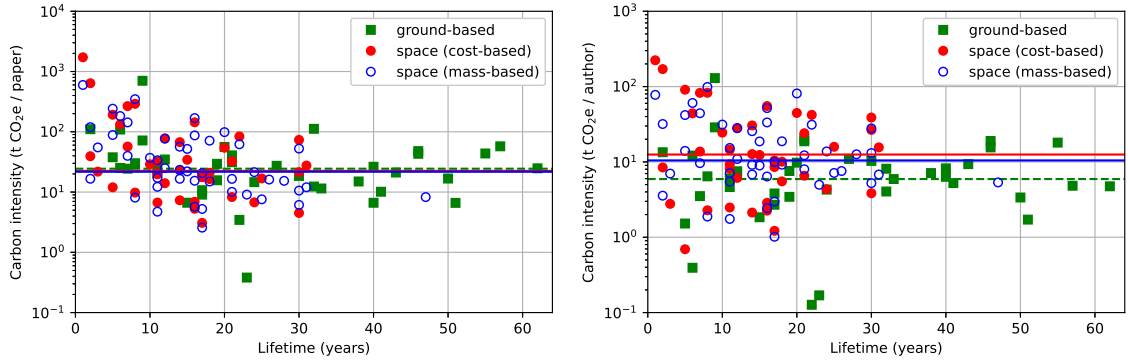}
\caption{
{\bf Carbon intensity versus time since launch or start of operations for space missions 
and ground-based observatories.} 
Open blue circles are for mass-based estimates and filled red circles are for cost-based estimates of 
space missions. 
Green squares represent ground-based observatories. 
The left panel shows the carbon intensity per peer-reviewed paper and the right panel shows the 
carbon intensity per unique author. 
The horizontal lines indicate the median carbon intensities.
\label{fig:ci}
}
\end{figure}

The results of this computation are shown in Tables \ref{tab:spacefootprint} and \ref{tab:groundfootprint}
and further illustrated in Fig.~\ref{fig:ci}, which shows the carbon intensities as function of a facility's lifetime. 
While there is an overall envelope of decreasing carbon intensities with time, not all facilities start with 
a large carbon intensity and some facilities still have an important carbon intensity decades after their launch 
or start of operations. 
Since the distribution of carbon intensities among the facilities is obviously heavily skewed, we 
use the median instead of the mean to estimate a typical value for the facilities. 
For space missions we find a median carbon intensity of 22 \tcoe\ per paper for both the mass-based 
and cost-based estimates. 
Ground-based observatories have a comparable median carbon intensity of 24 \tcoe\ per paper. 
When normalized by author, space missions have a median carbon intensity of 10 \tcoe\ per author for the 
mass-based and 13 \tcoe\ per author for the cost-based estimates, while ground-based observatories have 
a median value of 6 \tcoe\ per author. 
This suggests that, on average, more authors are involved in publications citing ground-based observatories 
compared with publications citing space missions.

To estimate the share of the carbon footprint that should be attributed to an astronomer at our institute, 
we multiply the annual carbon footprint of each facility with the fraction of unique authors of peer-reviewed 
papers published in 2019 that are affiliated to IRAP 
(alternative attribution methods are explored in the Supplementary Information). 
This results in a carbon footprint of 2.5$\pm$0.5 \ktcoe\ (mass-based) and 2.8$\pm$0.6 \ktcoe\ 
(cost-based) for the space missions and 1.3$\pm$0.5 \ktcoe\ for the ground-based observatories, 
totalling 4.0$\pm$0.7 \ktcoe\ for IRAP in 2019, where the uncertainty includes the difference 
between the mass- and cost-based estimates. 
For each of the 144 IRAP astronomers with a PhD degree this corresponds to a footprint of 27.4$\pm$4.8 \tcoe,
while for each of the 263 persons working at IRAP in 2019 (that is, including students and all technical and administrative 
staff) the footprint is 15.0$\pm$2.6 \tcoe. 
The research infrastructures listed in Tables \ref{tab:spacefootprint} and \ref{tab:groundfootprint} are 
however only a subset of all infrastructures that exist worldwide, hence formally these estimates are 
lower limits. 
Correcting for the incompleteness of our subset suggests a footprint of about 36 \tcoe\ per 
IRAP astronomer in 2019 (Supplementary Information).

\begin{figure}
\centering
\includegraphics[width=15cm]{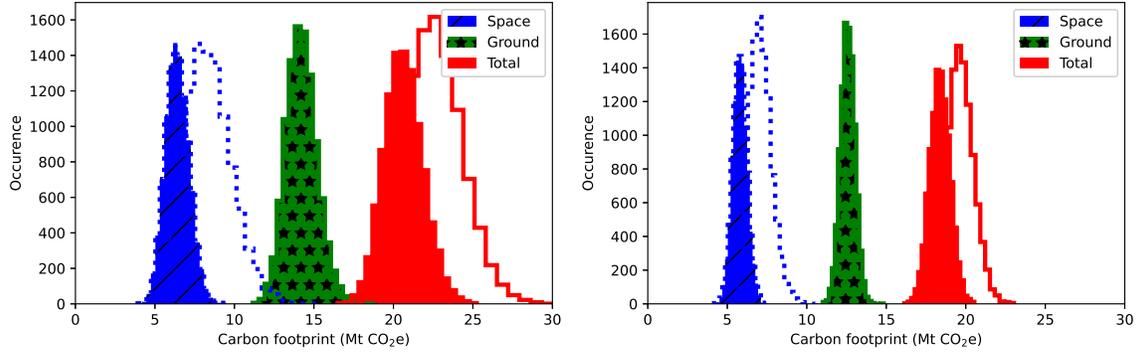}
\caption{
{\bf Distribution of the carbon footprint of astronomical research infrastructures that exist worldwide
as derived using the bootstrap method.}
The left panel shows the sampling of all facilities, and for the right panel the infrastructure 
with the largest carbon footprint in each category was excluded from the sampling. 
For space missions, cost-based estimates are shown as filled histograms and mass-based estimates 
are shown as unfilled histograms. 
Estimates for space missions are in blue hatched fill and dotted line, estimates for ground-based observatories 
are in green fill with stars and total carbon footprint estimates, which is the sum of space and ground-based 
facilities, are in red uniform fill and solid line.
\label{fig:bootstrap}
}
\end{figure}

Based on the results of Tables \ref{tab:spacefootprint} and \ref{tab:groundfootprint} and using an 
estimate for the total number of astronomical research infrastructures that were active in 2019 in 
the world, we estimate their global carbon footprint using a bootstrap method (see `The worldwide footprint of astronomical facilities' in Methods). 
While this method assumes that the research infrastructures considered in this paper are representative 
of all astronomical facilities that exist worldwide, we reduce the bias introduced by the 
specific selection by dividing all research infrastructures into broad categories that reflect 
scientific topic or observatory type. 
The results of this exercise are summarized in Table \ref{tab:worldfootprint} and the carbon footprint 
distributions that were obtained by the bootstrapping are shown in Fig.~\ref{fig:bootstrap}.

\begin{table}
\scriptsize
\caption{Extrapolated carbon footprint of all active astronomical research infrastructures in the world.
The number of active facilities in Tables \ref{tab:spacefootprint} and \ref{tab:groundfootprint} in a given 
category are given in column 2, the estimated worldwide number in each category is given in column 3.
The sum of the total and annual carbon footprints in each category for the selected infrastructures is 
given in columns 4 and 5, quoted uncertainties reflect the 80\% uncertainty in the footprint of individual 
facilities.
The extrapolated values for worldwide active research infrastructures are given in columns 6 and 7,
quoted uncertainties reflect the variance of the bootstrap sampling and the uncertainty of the carbon
footprint estimates.
\label{tab:worldfootprint}}
\centering
\vspace{6pt}
\begin{tabular}{l c c c c c c c c}
\hline
Category & \multicolumn{2}{c}{Research Infrastructures} && \multicolumn{2}{c}{Carbon footprint of selected} && \multicolumn{2}{c}{Carbon footprint of all} \\
& \multicolumn{2}{c}{} && \multicolumn{2}{c}{research infrastructures} && \multicolumn{2}{c}{research infrastructures} \\
\cline{2-3} \cline{5-6} \cline{8-9}
& Selected & Worldwide && Life cycle & Annual && Life cycle & Annual \\
& & && (\ktcoe) & (\ktcoeyr) && (\ktcoe) & (\ktcoeyr) \\
\hline
\multicolumn{7}{l}{Space missions (mass-based)} \\
Solar	 & 3 & 3 && 282$\pm$147 & 23$\pm$13 && 282$\pm$166 & 23$\pm$16 \\
Plasma & 7 & 13 && 553$\pm$219 & 31$\pm$12 && 1,029$\pm$375 & 58$\pm$20 \\
Planetary & 6 & 21 && 703$\pm$256 & 65$\pm$25 && 2,461$\pm$546 & 226$\pm$54 \\
Astro	 & 12 & 18 && 1,759$\pm$586 & 99$\pm$28 && 2,636$\pm$955 & 149$\pm$43 \\
Sum & & && 3,298$\pm$692 & 217$\pm$78 && 6,409$\pm$1,174 & 455$\pm$74 \\
\multicolumn{7}{l}{Space missions (cost-based)} \\
Solar	 & 3 & 3 && 510$\pm$241 & 39$\pm$19 && 510$\pm$247 & 39$\pm$20 \\
Plasma & 4 & 13 && 402$\pm$175 & 30$\pm$14 && 1,306$\pm$352 & 96$\pm$30 \\
Planetary & 6 & 21 && 886$\pm$351 & 83$\pm$34 && 3,100$\pm$798 & 289$\pm$80 \\
Astro & 12 & 18 && 2,339$\pm$1,033 & 114$\pm$41 && 3,526$\pm$1,816 & 172$\pm$68 \\
Sum & & && 4,137$\pm$1,131 & 265$\pm$58 && 8,442$\pm$2,030 & 596$\pm$111 \\
\multicolumn{7}{l}{Ground-based observatories} \\
OIR ($\ge3$m) & 9 & 37 && 1,037$\pm$478 & 42$\pm$21 && 4,263$\pm$665 & 171$\pm$27 \\
OIR (others) & 19 & $\sim$1000 && 65$\pm$17 & 3$\pm$1 && 3,409$\pm$163 & 147$\pm$6 \\
Radio & 6 & 74 && 206$\pm$81 & 10$\pm$4 && 2,540$\pm$345 & 127$\pm$18 \\
Radio Arrays & 9 & 27 && 1,156$\pm$481 & 87$\pm$46 && 3,465$\pm$1,140 & 260$\pm$115 \\
Others & 4 & 4 && 537$\pm$374 & 52$\pm$39 && 538$\pm$499 & 52$\pm$53 \\
Sum & & && 3,001$\pm$779 & 194$\pm$64 && 14,214$\pm$1,461 & 757$\pm$131 \\
\hline
\end{tabular}
\end{table}

We find that the world's fleet of astronomical space missions has a total carbon footprint of 
7.4$\pm$2.2 \Mtcoe\ and an annual footprint of 525$\pm$184 \ktcoeyr, where the uncertainties include 
the difference between the mass-based and cost-based estimates. 
We note that roughly half of the existing space missions are covered by Table \ref{tab:spacefootprint},
with a coverage of 100\% for solar missions, between 50\%--60\% for plasma or astrophysics 
missions and $\sim$30\% for planetary missions. 
Due to the strong representation of IRAP in many recent space missions, we thus consider 
that our extrapolation to the full inventory of active space missions worldwide is robust. 
We estimate the total carbon footprint of all ground-based astronomical observatories as 14.2$\pm$1.5 \Mtcoe\ and 
their annual footprint to be 757$\pm$131 \ktcoeyr. 
While the total footprint for ground-based observatories is about twice as large as that for space missions, 
their annual footprint is only about 44\% larger, owing to, on average, longer lifetimes for ground-based observatories 
compared with space missions.

Summing both contributions yields a total carbon footprint of all active astronomical research infrastructures 
in 2019 of 21.6$\pm$3.2 \Mtcoe\ and an annual footprint of 1,283$\pm$232 \ktcoeyr, where 
uncertainties reflect statistical variations introduced by the bootstrap method, the 80\% uncertainty in the 
emission factors for individual facilities and the difference between mass-based and cost-based estimates. 
By dividing the annual carbon footprint by an assumed world population of 30,000 astronomers with a PhD degree 
(see `The number of astronomers in the world' in Methods), we obtain an annual footprint of 42.8$\pm$7.7 \tcoeyr\ per astronomer. 
Assuming that astronomical institutes worldwide have similar staffing profiles as IRAP, which is roughly one 
student, technical or administrative personnel per astronomer with a PhD degree, the annual footprint per 
person working in astronomy in any capacity is roughly half this value.

We reiterate that these are order-of-magnitude estimates, obtained under the assumption that the 
facilities in Tables \ref{tab:spacefootprint} and \ref{tab:groundfootprint} are representative 
of all astronomical research infrastructures that exist worldwide. 
Using a modified procedure that avoids resampling of research infrastructures that have no equivalent 
in the world, such as the HST or the Atacama Large Millimeter/submillimeter Array (ALMA), results in a reduced total carbon footprint 
of 19.1$\pm$2.3 \Mtcoe\ and annual carbon footprints of 1,054$\pm$138 \ktcoeyr\ and 35.1$\pm$4.6 \tcoeyr\ 
per astronomer. 
Sampling of unique research infrastructures may hence overestimate the annual carbon footprint by 
about 20\%. 
Averaging the original and the modified bootstrap results and including their difference in the uncertainty 
yields a total carbon footprint of astronomical research infrastructures of 20.3$\pm$3.3 \Mtcoe\ and an annual 
carbon footprint of 1,169$\pm$249 \ktcoeyr\ and 39.0$\pm$8.3 \tcoeyr\ per astronomer.

The two methods we applied to derive the average carbon footprint of research infrastructures per 
astronomer holding a PhD, either based on more restrictive criteria for IRAP or based on an extrapolation 
to all active infrastructures for an average astronomer, lead to the same order-of-magnitude estimate. 
The range of values spanned by both estimates is 36.6$\pm$14.0 \tcoeyr, where the 
uncertainty mainly reflects the differences that are plausibly explained by the method of estimation. 
While the IRAP result of 27.4$\pm$4.8 \tcoeyr\ may miss the attribution of the carbon footprint of some 
infrastructures that were not considered in this paper, the global extrapolation of 
42.8$\pm$7.7 \tcoeyr\ using the original bootstrap method may overestimate the footprint due to the large 
footprints of a few facilities that are unique in the world.

Comparing our result of 36.6$\pm$14.0 \tcoeyr\ per astronomer for research infrastructures with estimates 
for other sources of GHG emissions due to astronomy research activity reveals that astronomical 
research infrastructures dominate the carbon footprint of an average astronomer. 
For example, excluding research infrastructures, capital goods and the purchase of goods and services, 
ref.~12 estimated the average carbon footprint of an astronomer at the Max Planck Institut 
for Astronomy in Heidelberg to be 18.1 \tcoeyr, dominated by 8.5 \tcoeyr\ from flights, 5.2 \tcoeyr\ from 
electricity use and 3.0 \tcoeyr\ from heating. 
For the average Australian astronomer (including PhD students), ref.~6 estimated a footprint 
of $\ge$37 \tcoeyr, comprised of 22 \tcoeyr\ for supercomputing, 6 \tcoeyr\ from flights, 4 \tcoeyr\ 
for powering the office buildings and 5 \tcoeyr\ for powering ground-based observatories in Australia. 
All of these individual contributions are below our estimate for the contribution of astronomical research 
infrastructures. 
While we cannot exclude that items such as the amortization of buildings and equipment or the purchase 
of goods and services that have generally not been quantified in carbon footprint assessments of 
astronomy institutes so far may yet represent a notable contribution in some cases, we know that their 
contribution at IRAP is substantially inferior to that of astronomical research infrastructures. 
We thus conclude that among the contributions that have so far been considered in the literature, space 
missions and ground-based observatories are the single most important contribution to the carbon footprint 
of an average astronomer.

\section*{Discussion}

It is clear from our analysis that the environmental sustainability of astronomical research 
infrastructures cannot be neglected if our community is seriously committed to reducing its 
overall carbon footprint. 
According to ref.~18, the Earth's land and oceans provide a carbon sink of about 
16 GtCO$_2$ yr$^{-1}$ (corrected for emissions from land-use change), corresponding to a per-capita 
carbon sink of 2 tCO$_2$ yr$^{-1}$. 
To reach net-zero GHG emissions, the per-capita emissions need to be reduced to that level. 
How the burden of emission reductions should be shared across humanity is a political question, for 
which consensus needs to be established collectively. 
Here we take a common global per-capita target, that is, the same emission target for all humans regardless 
of their nationality or country of residence, as a reasonable basis for initiating discussion. 
This would imply that an astronomer's footprint (including all their work-related activities, as well as 
their activities outside of work and lifestyle) must be reduced within the next decades to 2 tCO$_2$ yr$^{-1}$. 
In this scenario, emissions of research infrastructures would need to be divided by at least a factor of 20. 
In the interests of equity, policymakers may instead adopt nation- or region-specific emission targets. 
For example, to reach net-zero GHG emissions in France, Italy or Spain, consumption-based emissions 
would need to be reduced by about a factor of five and one could apply the same reduction factor to 
astronomers based in those countries. 
Australia, Canada and the United States currently have larger per-capita emissions and in those countries 
the reduction factors to reach net-zero GHG emissions would be closer to ten. 
A further proposition under consideration (for example, ref.~19) is that reduction factors should be based 
on national responsibility for climate breakdown. 
In this scenario, many countries in the global north would be required to close their research infrastructures, 
while countries in the global south would retain more capacity to develop new infrastructures. 
Whatever strategy for reduction factors is ultimately adopted, however, our results indicate that the required 
reductions will not be minor changes at the margin. 
Instead, they will fundamentally change how we do astronomy in the future. 
Astronomy is, of course, not the only scientific domain that heavily relies on research 
infrastructures and it is likely that similar conclusions will be reached in other research fields.

A necessary first step is that every existing and planned facility should conduct a comprehensive environmental 
life-cycle assessment and make the results public. 
The International Astronomical Union (IAU) could, for example, hold a register of all environmental life-cycle 
assessments, which would constitute a comprehensive database of knowledge for further studies. 
Funding agencies and space agencies have a critical role in ensuring that such analyses are conducted and 
published for each facility that they support. 
For example, based on our emission factors, the James Webb Space Telescope would have a carbon footprint 
between 310 \ktcoe\ (mass-based) and 1,223 \ktcoe\ (cost-based) and for the Square Kilometre Array we estimate the construction 
footprint to be 312 \ktcoe\ and the annual operations footprint to be 18 \ktcoeyr.
These figures are comparable with the largest footprint estimates in our study. 
It is urgent that we consolidate these estimates and implement effective measures to reduce the parts of their 
carbon footprint over which we still have control.

For existing infrastructures, the analysis should be used to prepare an action plan for how to reduce the 
footprint over the coming years. 
The footprint reductions should be monitored and regularly checked against the action plans and the plans 
should be adapted if needed. 
One could mimic the scheme of nationally determined contributions that was agreed upon in the 21st 
Conference of the Parties Paris Agreement and hold a register of all action plans and achievements at IAU that is publicly 
accessible. 
While compliance with carbon footprint reductions cannot be enforced on observing facilities, publishing 
the action plans and achievements would at least guarantee a transparent and open discussion in the 
community.

For planned infrastructures, the environmental life-cycle assessments should inform the decision about 
implementation. 
Abandoning future projects on the basis of their unacceptably high carbon footprint should be an option, 
but we emphasize that informed decisions of this nature require robust estimates. 
Having a centralized inventory of environmental life-cycle assessments of all existing infrastructures would 
allow the community to determine whether there is a margin in the sustainable carbon budget for new 
infrastructures. 
In particular, funding agencies and space agencies should include carbon budget limits in their 
roadmaps for future research infrastructures, assuring their compliance with the boundaries of our planet.

It is questionable whether the required reduction of the carbon footprint of research infrastructures can 
be reached within the next few decades using the measures described above, in particular if new 
infrastructures continue to be proposed and developed at the current pace. 
A possible solution that is often mentioned in the context of carbon footprint reductions is offsetting, 
yet effective offsetting requires substantial investments that are commensurable with the cost 
of building an astronomical research infrastructure (Supplementary Information). 
For existing facilities, the focus must obviously be put on their decarbonization, and if this is not 
sufficient then we must face the question of which infrastructures should be kept open and which 
should be shut down.

For planned new facilities, we must recognize that infrastructure investments that are made today lock 
in their carbon footprint over decades and that replacing fossil fuel for hardware, propellant and electricity 
production, transport and space launches with renewable energies requires time and investment and 
probably new technologies that do not yet exist. 
Global warming is a rate problem: it is the amount of \co\ we emit each year that is too high relative 
to the environment's capacity to absorb and to recycle, producing excess atmospheric \co\ 
concentrations that heat our planet. 
Spreading the roughly 35 Gt\co\ that humanity emits every year\cite{friedlingstein2020} over a 
longer period (for example, 3 years) would bring the per-capita footprint down to about 2 tCO$_2$ yr$^{-1}$.
Making the annual footprint of astronomical research infrastructures of 1,169 \ktcoeyr\ compliant with 
60 \ktcoeyr, which is the equal-share target for 30,000 astronomers, implies spreading the annual 
footprint over about 20 years.

We therefore believe that reducing the pace at which we build new astronomical research 
infrastructures is the only measure that can make our field sustainable in the short run. 
This does not mean that we must stop developing new observatories or space missions, but we must 
do so at a (considerably) slower rate. 
Once the economy is substantially decarbonized, the rate of construction of new research infrastructures 
could be increased, considering their potential impact on climate change as well as other 
detrimental effects on the environment, including biodiversity loss, mineral extraction and use of water 
resources. 
The good news is that there is no imperative in science that fixes the rate by which new research 
infrastructures need to be constructed. 
Today, the rate is determined by our imagination and, ultimately, by money. 
Tomorrow, it must be determined by sustainability.

Reducing the pace has many side benefits, some of which have already been recognized earlier 
by the Slow Science movement\cite{alleva2006}: more comprehensive exploitation of data, 
more time for in-depth science, less publication pressure and more money available to move the 
already existing infrastructures towards sustainability. 
The solution is in our hands, the only question is whether the astronomical community will choose to 
recognize and make use of it.

\vspace{12pt}
\begin{methods}

\subsection{Carbon footprint estimation}

To estimate the carbon footprint of astronomical research infrastructures, we follow the method developed 
by the French Agency for Ecological Transition (ADEME; \url{https://www.bilans-ges.ademe.fr/en/accueil}) and 
the French Association Bilan Carbone (ABC; \url{https://www.associationbilancarbone.fr}). 
This method includes the definition of boundaries for the exercise, the collection and analysis of the 
relevant data and the proposal of an action plan for the reduction of the carbon footprint.

In this method, the carbon footprint of a research infrastructure is defined as an aggregate 
of all GHG emissions that are generated within the defined boundaries. 
To enable aggregation of the different gases that cause global warming, GHG emissions are 
converted into amounts of carbon dioxide equivalents, denoted \coe, which take into account 
the different warming potentials and lifetimes of the various GHGs emitted into the atmosphere. 
Specifically, the carbon footprint is computed using
\begin{equation}
\mbox{Carbon footprint} = \sum_i A_i \times EF_i
\label{eq:footprint}
\end{equation}
where $A_i$ are called activity data (for example, MWh of electricity consumed, tonnes of concrete poured 
or \meuro\ of money spent), 
$EF_i$ are called emission factors (for example, \tcoe\ emitted per MWh consumed, \tcoe\ emitted per 
tonne of concrete poured or \tcoe\ emitted per \meuro\ of money spent) 
and the sum is taken over all relevant activities that have been identified within the defined 
boundaries.

Since activity data for space missions or ground-based observatories are scarce, we will primarily 
use an EIO analysis, which in our case amounts to using the cost of a space 
mission, or the cost for construction and operations of a ground-based observatory, as the relevant activity 
data. 
For space missions, we cross-check our analysis by adopting a second approach, which uses the 
satellite payload launch (or wet) mass as activity data. 
This means that for space missions equation (\ref{eq:footprint}) simplifies to
\begin{equation}
\mbox{Carbon footprint}_{\rm c/m} = A_{\rm c/m} \times EF_{\rm c/m}
\end{equation}
where $A_{\rm c/m}$ is either mission cost or payload launch mass and $EF_{\rm c/m}$ the corresponding 
emission factor, while for ground-based observatories we use
\begin{equation}
\mbox{Carbon footprint} = A_{\rm co} \times EF_{\rm co} + A_{\rm op} \times T \times EF_{\rm op}
\end{equation}
where $A_{\rm co}$ is the construction cost, $A_{\rm op}$ the annual operating cost, $T$ the lifetime
in years and $EF_{\rm co}$ and $EF_{\rm op}$ are the emission factors for construction and annual operations, 
respectively. 
For many activities, emission factors can, for example, be found in ref.~15, yet such databases 
do not contain specific emission factors for `\tcoe\ per kg of payload mass' or `\tcoe\ per 
\meuro\ of operating costs'. 
We therefore estimate dedicated emission factors based on published carbon footprint 
assessments for astronomical research infrastructures for our analysis (see `Emission factors').

We aim for using activity data and emission factors that cover scopes 1--3 of a carbon 
footprint assessment, where scope 1 refers to direct emissions from sources owned or controlled by a 
research infrastructure (that is, any owned or controlled activities that release emissions straight into the atmosphere, 
such as a gas or diesel generator owned by an observatory), scope 2 refers to indirect emissions from the 
consumption of purchased electricity, heat, steam or cooling (for example, emissions related to purchased electricity 
needed for observatory operations) and scope 3 refers to indirect emissions from other activities not directly 
controlled by the research infrastructure (for example, emissions related to the construction phase of a facility and 
employee travel). 
We approach this goal by using emission factors that are based on scope 1--3 analyses, yet 
we recognized that in some cases the activity data used do not fully cover all scopes and consequently 
our carbon footprint assessment presents therefore only a lower limit to a full scope 1--3 analysis.

We furthermore aim for a full LCA, covering 
phase A (feasibility), 
phase B (preliminary definition), 
phase C (detailed definition), 
phase D (qualification and production), 
phase E (utilization) and 
phase F (disposal) 
for space missions and design and definition, construction, operations and dismantling 
for ground-based observatories. 
Yet the feasibility and disposal phases are generally excluded from the activity data, which also do not 
cover the long-term conservation of the acquired data after the end of a space mission. 
We also note that for space missions, information on mission extension was not always available and 
consequently the carbon footprint assessment may be limited to the initial mission definition, which often 
covers only the first few years of operations. 
For ground-based observatories the footprint is after a few decades dominated by the operations 
footprint, however the full operational lifetime of an observatory, and hence its life-cycle footprint,
is not known in advance. 
We therefore decided to limit the operations part of our carbon footprint assessment to the end of our 
reference year 2019.

As this work was conducted in the context of our institute's effort to estimate its carbon footprint, we 
considered all space missions and ground-based observatories that were used to produce peer-reviewed 
journal articles authored or co-authored by IRAP scientists in 2019. 
To achieve this, we scrutinized the titles and abstracts of all publications signed by authors that were 
affiliated to IRAP in 2019 using the Astrophysics Data System (ADS) and collected a list of all research 
infrastructures from which data were used to produce the papers. 
In a few cases where the usage of research infrastructures was not obvious from the title or abstract of 
the paper, we also scrutinized the full text of the publication. 
In total, we identified 46 space missions and 39 ground-based observatories that were used by 
IRAP scientists to produce peer-reviewed journal articles published in 2019.

\subsection{Emission factors}

\subsubsection*{Space missions}

There exist no publicly available data on the life-cycle carbon footprint of planned or existing space 
missions, despite the work that has been done in this field for more than a decade or so (for example, by ESA 
in the context of the Clean Space initiative)\cite{austin2015}. 
ESA maintains a comprehensive LCA database yet declined to share 
it with us for our work. 
But even if ESA had provided its database, without detailed activity data the database would 
have been of very limited use for our work.

Nevertheless, ESA has published relative contributions of the life-cycle breakdown for some launchers 
and space missions, which allows us to perform consistency checks and order-of-magnitude estimates. 
For example, for the Sentinel-3B mission, an Earth observation satellite of the EU Copernicus programme, 
ref.~23 found that 44\% of the carbon footprint is attributable to launcher-related activities, 25\% to 
the operations phase and 19\% to the definition, qualification and production of the spacecraft, referred 
to as phase C+D. 
In their life-cycle assessment of the Astra-1N and the MetOP-A missions, ref.~24 also find that 
the launch campaign dominates the carbon footprint (59--64\%), followed by phase C+D (27--28\%), 
manufacturing, assembly, integration and test (7--10\%) and operations (1--2\%). 
Overall, it is estimated that between 50\% to 70\% of the carbon footprint of a space mission is related to the 
launcher production, launch campaign and launch event, depending on the launcher's dry mass\cite{maury2019}.

In our work, we will use an estimate for the life-cycle carbon footprint of a space mission 
that is based on a space-specific life-cycle sustainability assessment (LCSA) framework developed 
by ref.~16 during his PhD thesis and that he carefully cross-checked with the ESA tools and 
database. 
He applied his framework for two case studies: the M\`IOS mission, a small satellite mission to 
the Moon that aimed to collect data on the micrometeorite and radiation environment and detect the presence 
of water/ice on the lunar south pole in view of a future Moon base, and the NEACORE mission, a set of 
six nanosatellites for the exploration of asteroids by collision and flyby reconnaissance.

In the ref.~16 case study, the M\`IOS mission is launched with Ariane 5 ECA together with 3 
other missions, hence only 25\% of the carbon footprint of the launch segment is attributed to the 
M\`IOS mission. 
The payload has a wet mass of 286 kg and the assumed mission duration is 2 years. 
Reference 16 estimated the total carbon footprint of the mission to be 11,200 \tcoe\ and a 
cost of \euro 165 million after applying an inflation correction to convert to 2019 economic conditions, 
resulting in a monetary emission factor of 68 \moneyeft\ and a per payload wet-mass emission factor of 39 \masseft.

The NEACORE mission is launched with a dedicated PSLV-CA rocket. 
The 6 nanosatellites have a total wet mass of 143 kg and the assumed mission duration is 4 years and 8 months. 
Reference 16 estimated the total carbon footprint of the mission to be 8,780 \tcoe\ and a 
cost of \euro 41 million when converted to 2019 economic conditions, resulting in a monetary emission factor of 
214 \moneyeft\ and a per payload wet-mass emission factor of 61 \masseft.

Taking the mean value of the results from ref.~16 leads to emission factors of 140 \moneyeft\ and 
50 \masseft, which are the values that we adopt for our study. 
We note that the M\`IOS and NEACORE estimates differ by less than 53\% from these mean values, which 
is well within the 80\% uncertainty that we adopt for individual facilities in our study.

\subsubsection*{Ground-based observatories}

For ground-based observatories, we use monetary emission factors that we estimate separately for 
construction and operations since the related activities may have different carbon footprint breakdowns. 
For example, construction often needs important quantities of concrete and steel, with a very high carbon 
footprint of 1,700--1,800 \moneyeft, while operations typically consume a lot of electricity, which, for 
example, in Chile has a typical carbon footprint of 2,500 \moneyeft\ (ref.~15). 
We did not apply a similar separation in our analysis of space missions, since their operations footprint 
is estimated to be small compared with their construction footprint\cite{chanoine2017, desantis2013}.

\paragraph{Construction}

Reference 26 has estimated the carbon footprint of the Giant Radio Array for Neutrino Detection 
(GRAND), an experiment that aims to detect ultra-high-energy neutrinos with an array of radio 
antennas. 
The ref.~26 carbon footprint estimate for GRAND includes travel, digital technologies and the 
hardware equipment. 
For the 3 stages of the project, lasting 25 years in total, the authors obtained a total carbon footprint 
of 147,220 \tcoe. 
The footprint is dominated by the production of stainless steel for the antennas, data storage and data transfer. 
Excluding the footprint of digital technologies (as it is mostly related to operations and not construction), 
results in a construction footprint of 81,239 \tcoe. 
A construction cost estimate of 200 \meuro\ for the GRAND project has been quoted by ref.~27, excluding the 
cost of rent and salaries. 
Using that cost estimate results in a formal monetary emission factor of 406 \moneyeft\ for GRAND 
construction, which is a plausible order-of-magnitude estimate that is in close agreement with the monetary 
emission factor for electrical and information technology equipment (400 \moneyeft)\cite{basecarbone}.

For the European Extremely Large Telescope (E-ELT), a 39.3-m-diameter optical and near-infrared 
telescope that is currently under construction in Chile, the European Southern Observatory estimates 
the construction carbon footprint to be 63.7 \ktcoe, covering different quantities of material as well as the 
energy needed for some parts of the work (for example, blasting, road construction and so on) (ESO, personal communication). 
Using the E-ELT construction cost of 945 \meuro, quoted in the E-ELT construction proposal\cite{eltproposal}
(converted to 2019 economic conditions), results in a monetary emission factor of 
67 \moneyeft, which is considerably smaller than the GRAND estimate and considerably lower than any 
sector-based estimate in ref.~15.

Since the E-ELT estimate does not cover all construction activities, the derived emission factor presents what is clearly a 
lower limit, while with a cost estimate that does not comprise the full project cost, the GRAND estimate 
presents what is clearly an upper limit to the emission factor. 
In the absence of more reliable information for the construction-related footprint for ground-based observatories, 
we take the mean value of 240 \moneyeft\ of the GRAND and E-ELT estimates and note that the individual 
estimates are within our adopted uncertainty of 80\% of this mean value.

\paragraph{Operations}

According to ESO (personal communication), the carbon footprint of the Paranal and the La Silla observatories in 
2018 were 8.6 \ktcoeyr\ and 2.3 \ktcoeyr, respectively. 
Both footprints are dominated by energy use (71\% for Paranal, 92\% for La Silla), followed 
by commuting (11\% for Paranal, 5\% for La Silla) and capital goods (15\% for Paranal, 2\% for La 
Silla). 
Some of the purchases for Paranal and La Silla are accounted for in the carbon footprint estimated 
by ESO (personal communication) for the Vitacura site and following the recommendation by R. Arsenault (personal communication) 
we added 30\% of the purchase footprint of the Vitacura site to both the Paranal and La Silla footprints, 
resulting in 9.4 \ktcoeyr\ for Paranal and 2.8 \ktcoeyr\ for La Silla.

The annual operations budget of the Paranal observatory in 2011 has been quoted as 16.9 \meuro\ (18.6 \meuro\
in 2019 economic conditions) plus 174 full-time equivalents (a measurement of workforce 
employed as equivalent to full-time employees)\cite{paranal}. 
Assuming 120 \keuro\ per full-time equivalent (2019 economic conditions; ref.~30), one can estimate the 
annual operations budget of Paranal in 2019 to be 39.5 \meuro, which results in a monetary emission 
factor of 238 \moneyeft\ for Paranal operations. 
For La Silla, ref.~31 has quoted an operations budget of 5.9 \meuro\ in 2004 (7.9 \meuro\ in 
2019 economic conditions), resulting in a monetary emission factor of 354 \moneyeft\ for La Silla 
operations.

For the Canada--France--Hawaii Telescope (CFHT), ref.~32 has estimated the 2019 GHG emission 
of CFHT operations, including travel, the CFHT-owned vehicle fleet and on-site energy consumption. 
The total carbon footprint for these items amounts to 749 \tcoe, dominated by power generation (63\%)
and transportation (31\%). 
For an operations budget of 6.3 \meuro\ for the same year\cite{cfht2019}, this results in a monetary 
emission factor of 119 \moneyeft\ for CFHT operations, considerably below the estimate for Paranal 
and La Silla. 
The authors note that flight-related emissions may be underestimated in their study and alternative 
estimators for flight-related emissions\cite{barret2020} indeed have yielded an estimate that is higher by a 
factor of two. 
Doubling the contribution of flights in the original calculation by ref.~32 would result 
in a total footprint of 923 \tcoe\ and a monetary emission factor of 148 \moneyeft\ for CFHT operations. 
Additional emission sources such as the purchase of materials or food for employees that were not 
considered by ref.~32 would further increase this factor.

Based on these 3 estimates, we adopt an average monetary emission factor of 250 \moneyeft\ for 
observatory operations in our analysis, which is the rounded mean value of the Paranal, La Silla and 
CFHT estimates. 
All specific estimates above are well within our adopted 80\% uncertainty of this average. 
We note that the exact value of this factor is strongly dependent on the carbon intensity of the means 
of electricity production that is used to power the observatory. 
The same applies to computing and data storage, as well as ground support for space 
missions. 
More dedicated life-cycle assessments of specific research infrastructures are needed here to better 
understand the importance of these contributions to the total carbon footprint.

\paragraph{Adopted emission factors}

The emission factors that we use throughout this study are summarized in Table \ref{tab:emission-factors}. 
For comparison, typical monetary emission factors in France range from 110 \moneyeft\ for insurance 
and banking, over 400 \moneyeft\ for electrical and optical information technology and office equipment, 700 \moneyeft\ for 
machinery equipment, up to 1,700 \moneyeft\ for metals and 1,800 \moneyeft\ for mineral products\cite{basecarbone}. 
The relatively low monetary emission factor for space missions is related to the fact that space missions 
are much less material intensive compared with ground-based observatories after normalizing by cost. 
For example, the liftoff mass of a \euro 1 billion space mission launched with Ariane 5 ECA is about 
790 tonnes\cite{ariane}, while the E-ELT (which has a similar cost) has a mass of about 60,000 tonnes\cite{eltweb}. 
The space sector is in fact unique, and is characterized by low production rates, long development 
cycles and specialized materials and processes\cite{geerken2018}.

We also note that the emission factor for ground-based observatory construction is based on rather 
recent estimates, while some of the observatories that we consider were constructed several decades ago, 
at a time when the carbon intensity of construction was probably larger. 
Our construction footprint estimates for the corresponding observatories are therefore likely to be lower limits.

We stress that the values of the emission factors that we adopt remain rather uncertain, given the 
scarcity of publicly available information relating to life-cycle assessments for astronomical research infrastructures. 
Reference 15 has quoted a typical uncertainty of 80\% for monetary emission factors, which corresponds 
to the typical scatter of the scarce data about the adopted emission factors in Table \ref{tab:emission-factors}.
This uncertainty clearly dominates the uncertainties of our results. 
We note, however, that our monetary emission factors are already on the low side of the sector-based 
estimates in ref.~15, hence it seems unlikely that our adopted values are notably 
overestimated. 
The true emission factors and the resulting carbon footprint could easily be larger. 
More work is needed in this area, and we urge the agencies responsible for existing and future infrastructures 
for astronomy research, as well their public and industry partners, to contribute actively to consolidating 
carbon footprint estimates.

\subsection{Annual footprint}

To compute the annual carbon footprint of a research infrastructure, we need to devise a method of 
how to distribute the life-cycle footprint of space missions or the construction footprint of ground-based 
observatories over the years. 
For that purpose, we define as the annual carbon footprint the total carbon footprint divided by the 
number of years over which the greenhouse gases were emitted. 
The accounting is started with the launch for a space mission or with start of operations for 
ground-based observatories. 
For space missions, we divide the lifetime carbon footprint by the lifetime of the mission. 
For missions that were launched recently, we assume a lifetime of 10 years. 
In other words, we assume a minimum operations period of 10 years for all infrastructures, avoiding 
attributing an artificially large annual GHG emission to missions that were launched recently. 
For ground-based observatories, we divide the construction footprint by the lifetime of the 
observatory and add the annual operations footprint. 
For observatories that were built after 2009, we assume a lifetime of at least 10 years, otherwise 
we take the number of years since first light as the lifetime.

\subsection{The user community of astronomical research infrastructures}

To put the carbon footprint results in perspective, we estimated for each research infrastructure the 
total number of peer-reviewed papers that either analyse data from a given infrastructure or that refer 
to analysed data from the infrastructure or to the infrastructure itself. 
We obtained the number of such publications from ADS, using a full-text search over the period 
from launch until our reference year of 2019. 
We constructed a dedicated query string for each infrastructure with the aim to cover as many 
infrastructure-related publications as possible while keeping the false positives at a 
minimum (the script that we used for this work, including the definition of our query strings, 
can be accessed at \url{https://doi.org/10.5281/zenodo.5835840}). 
We checked the latter by visually investigating a sample of $\sim$100 results for each query. 
We used the same ADS query to assess the number of unique authors that signed at 
least one of the peer-reviewed papers. 
While the number of papers can be considered as a measure of the infrastructure impact, 
the number of unique authors can be considered as a measure of the size of the community 
that is using the infrastructure.

\subsection{Carbon intensity}

Using the above estimates of the user community we can then compute the carbon intensity of 
a facility, which we define as the lifetime carbon footprint divided by either the number of peer-reviewed 
papers or the number of unique authors. 
The meaning of these quantities is that they relate the total carbon footprint of a given infrastructure 
to the scientific productivity of the community and to the size of the community that makes use of it. 
The units of the carbon intensity should, however, not lead to the conclusion that writing a paper will 
actually have the quoted carbon footprint or adding one more author will increase the carbon 
footprint. 
While writing a paper has some carbon footprint, we do not estimate its value here as it 
is not in the scope of this paper and it will be small compared with the per-paper footprint of 
research infrastructures\cite{song2015}. 
What we estimate is the share of the carbon footprint of a given infrastructure among the scientific 
community, and writing one more paper or adding one more author will actually reduce the carbon 
intensity as it will increase the denominator in the equations. 
This means that the carbon intensity is a dynamic quantity that will evolve with time 
(Fig.~\ref{fig:ci}).

\subsection{The worldwide footprint of astronomical facilities}

The research infrastructures that we selected for this study obviously represent only a subset of all 
astronomical research infrastructures that exist worldwide and reflect the specific research activities 
of the scientists at IRAP. 
We can, however, extrapolate the results to all astronomical research infrastructures to estimate 
their total carbon footprint globally. 
For this exercise, we limit ourselves to the research infrastructures that were active in 2019, disregarding 
the space missions in Table \ref{tab:spacefootprint} that had stopped operating. 
We divide all research infrastructures into broad categories that reflect scientific topic 
(for example, solar, plasma or astro) or observatory type to reduce the bias introduced by the specific 
selection of infrastructures in Tables \ref{tab:spacefootprint} and \ref{tab:groundfootprint}. 
We note that we split optical-near-infrared (OIR) telescopes into facilities with diameter of 3 m or 
larger and those with smaller diameters since they differ substantially in carbon footprint and in total 
number. 
While the total number of telescopes with mirrors of at least 3 m diameter is easy to determine, we only 
have an approximation for the number of smaller telescopes. 
For example, ref.~37 provides a directory of about 1,000 professional astronomical 
observatories and telescopes, while the IAU Minor Planet Center lists 2,297 observatory codes\cite{iau-obs}. 
We therefore make our estimate by assuming that there exist at least 1,000 professional OIR telescopes 
worldwide with diameters smaller than 3 m. 
Due to their vastly different footprints, we also separate single-dish radio telescopes from multiple-dish 
or antennae arrays. 
Finally, we also list all other instruments, including the Stratospheric Observatory for Infrared Astronomy (SOFIA), but do not extrapolate this number due to the 
heterogeneity of the infrastructures in this category.

For each of the categories, we determined the number of active research infrastructures in our list and 
compared them with the total number of research infrastructures that were active worldwide in 2019. 
The full list of facilities that we considered is provided in Supplementary Data 1. 
We then did a bootstrap sampling by randomly selecting $N$ times an infrastructure from the 
list of active infrastructures in a given category, where $N$ is the total number of research infrastructures 
that exist worldwide in a given category. 
We note that a given infrastructure in our list can be sampled several times in this process. 
Summing the carbon footprints for the $N$ sampled infrastructures then provides an estimate for the total 
carbon footprint of all infrastructures that exist worldwide. 
We repeated the random sampling 10,000 times and computed the mean and standard deviations of all 
estimates to assess the carbon footprint of all infrastructures and the uncertainty that arises from the 
random sampling procedure. 
The resulting distribution of carbon footprints is shown in the left panel of 
Fig.~\ref{fig:bootstrap}.

We reiterate that the results obtained by this method are order-of-magnitude estimates, obtained under the 
assumption that the research infrastructures in Tables \ref{tab:spacefootprint} and \ref{tab:groundfootprint} 
are representative for all astronomical research infrastructures that exist worldwide. 
To estimate the sensitivity of our result to this assumption, we repeated the bootstrap procedure by excluding the 
infrastructure with the largest carbon footprint in each category from the sampling, adding its footprint 
directly to the result. 
This avoids sampling infrastructures that are unique in terms of carbon footprint and for which no equivalent 
exists worldwide, such as, for example, the HST or the ALMA observatory, multiple times. 
This reduces the carbon footprint estimates and their sampling spread, as illustrated in the right panel of 
Fig.~\ref{fig:bootstrap}.

\subsection{The number of astronomers in the world}

To estimate how much research infrastructures contribute to the annual carbon footprint of an average 
astronomer, we need to estimate the number of professional astronomers worldwide. 
The IAU lists 12,165 members at the beginning of August 2021\cite{iau-members}, yet not every 
astronomer is a member of the IAU. 
Reference 40 studied the fraction of astronomers with PhD degrees who joined IAU for 12 countries 
(including astronomers, astrophysicists, physicists who study cosmology, high energy 
astrophysics or astroparticle physics), finding that on average 51\% of all astronomers with PhD 
degrees are members of the IAU. 
This would imply that there are 24,000 astronomers with PhD degrees worldwide. 
The country where the fewest astronomers joined the IAU is the United States, with only 40\% of astronomers 
being members of IAU. 
Applying this fraction would imply that there are 30,000 astronomers with PhD degrees worldwide, 
which probably is an upper limit. 
We apply this number in our work to make sure that we do not overestimate the per-astronomer 
footprint of astronomical research infrastructures.
\end{methods}

\section*{Data Availability}

All data used for this work are available for download at \url{https://doi.org/10.5281/zenodo.5835840}.

\section*{Code Availability}

All code used for this work is available for download at \url{https://doi.org/10.5281/zenodo.5835840}.

Received: 8 October 2021; Accepted: 21 January 2022; Published online: 21 March 2022


\begin{addendum}

\item[Acronyms]
Acronyms used throughout this paper represent:
Hubble Space Telescope (HST),
International Gamma-Ray Astrophysics Laboratory (INTEGRAL),
X-ray Multi-Mirror mission (XMM),
Rossi X-ray Timing Explorer (RXTE),
Solar Dynamics Observatory (SDO),
Mars Atmosphere and Volatile Evolution mission (MAVEN),
R\"ontgensatellit (ROSAT),
Mars Reconnaissance Orbiter (MRO),
Solar and Heliospheric Observatory (SoHO),
Astronomy Satellite (AstroSat),
Magnetospheric Multiscale mission (MMS),
Solar Terrestrial Relations Observatory (STEREO),
Advanced Composition Explorer (ACE),
Interior Exploration using Seismic Investigations, Geodesy and Heat Transport probe (InSight),
Parker Solar Probe (PSP),
Wide-field Infrared Survey Explorer (WISE),
Thermosphere, Ionosphere, Mesosphere Energetics and Dynamics mission (TIMED),
Interplanetary Monitoring Platform 8 (IMP-8),
Neutron star Interior Composition Explorer (NICER),
Nuclear Spectroscopic Telescope Array (NuSTAR),
Transiting Exoplanet Survey Satellite (TESS),
Galaxy Evolution Explorer (GALEX),
Detection of Electromagnetic Emissions Transmitted from Earthquake Regions satellite (DEMETER),
Very Large Telescope array (VLT),
Atacama Large Millimeter/submillimeter Array (ALMA),
Stratospheric Observatory for Infrared Astronomy (SOFIA),
Anglo-Australian Telescope (AAT),
Very Large Array (VLA),
Very Long Baseline Array (VLBA),
Institut de radioastronomie millim\'etrique (IRAM),
Canada France Hawaii Telescope (CFHT),
European Southern Observatory (ESO),
Green Bank Telescope (GBT),
Low-Frequency Array (LOFAR),
James Clerk Maxwell Telescope (JCMT),
Australia Telescope Compact Array (ATCA),
High Energy Stereoscopic System (H.E.S.S.),
More of Karoo Array Telescope (MeerKAT),
Gran Telescopio Canarias (GTC),
Nobeyama Radio Observatory (NRO),
Large Millimeter Telescope (LMT),
Mauna Loa Solar Observatory (MLSO),
Atacama Pathfinder Experiment (APEX),
Submillimeter Array (SMA),
Event Horizon Telescope (EHT),
Telescope Bernard Lyot (TBL),
Observatoire de Haute-Provence (OHP),
Korea Microlensing Telescope Network (KMTNet),
T\'elescope H\'eliographique pour l'Etude du Magn\'etisme et des Instabilit\'es Solaires (THEMIS),
Yunnan Astronomical Observatory (YAO),
Himalayan Chandra Telescope (HCT),
Fred Lawrence Whipple Observatory (FLWO),
National Astronomical Observatory San Pedro M\'artir (OAN-SPM),
Bohyunsan Optical Astronomy Observatory (BOAO),
Optical Gravitational Lensing Experiment (OGLE),
Centre P\'edagogique Plan\`ete Univers (C2PU),
T\'elescope \`a Action Rapide pour les Objets Transitoires (TAROT), and
Nanshan One meter Wide-field Telescope (NOWT).

\item
We thank R.~Arsenault for the insights that he provided on the carbon footprint 
estimates of ESO infrastructures and sites.
We furthermore thank
M.~de Naurois,
C.~Duran,
Z.~Fan, 
C.-U.~Lee,
A.~Klotz,
K.~Kotera,
K.~Tatematsu and
S.~O'Toole 
for having provided data that were useful for this research.
In addition we thank
N.~Flagey,
L.~Pagani,
G.~Song,
M.~Smith-Spanier and
A.~Ross Wilson
for useful discussions
and K.~Lockhart and S.~Blanco-Cuaresma for their help with ADS.
This research has made use of NASA's Astrophysics Data System Bibliographic Services.
In addition, this work has made use of the Python 2D plotting library matplotlib\cite{hunter2007}.

\item[Author Contributions Statement]

J.K. gathered the activity data, made the estimates of the emission factors, estimated the
carbon footprints and drafted the paper.
All authors defined the analysis method and the IRAP carbon footprint attribution method,
elaborated the discussion section and reviewed the manuscript.

\item[Correspondence] 
Correspondence should be addressed to J.K. (jurgen.knodlseder@irap.omp.eu).

\item[Competing Interests]
The authors declare no competing interests.

\item[Peer review information]
{\em Nature Astronomy} thanks Andrew Wilson, Robin Arsenault
and Lewis Ball for their contribution to the peer review of this work.

\newpage
\renewcommand\tablename{Supplementary Table}
\setcounter{table}{0}
\item[Supplementary Information]
\hfill \break
\section*{Activity data}

\subsection{Space missions}

We gathered full mission cost and payload launch mass estimates for all selected missions from the 
internet.
The results of this exercise are summarised in Supplementary Table \ref{tab:spaceactivity}.
Payload launch mass estimates are straightforward to find on the Wikipedia pages of the corresponding
space mission, from which we also gathered the launch and mission end dates. 
Reliable full mission cost estimates are significantly more difficult to compile, as mission-related
documentation is voluminous, and press articles rarely include cost breakdowns. 
Our mass-based estimates are therefore more homogenous, but they do not account for the complexity 
of a space mission that also drives its final cost and eventually the carbon footprint. 
Likewise, they do not reflect differences in mission duration that may lead to differences in carbon footprints
for the operations phase.
In general, the full mission cost includes the cost of in-kind contributions to the satellite payload, yet there 
may be exceptions where these costs are not included.
We did not explicitly account for the additional costs of mission extensions.  
We found that for some cases mission extensions were included in the quoted costs, in particular for all of
NASA's planetary missions where annual budgets are provided by
\cite{cost-planetary}.  
In other cases, the extensions were not included, and in a few cases, the construction cost value quotes 
we found were suspiciously low.

\begin{table}
\tiny
\caption{Payload launch mass and mission cost estimates for the space missions considered in this
work. For a few missions no cost estimates could be found, the corresponding entries in the tables
are blank.
\label{tab:spaceactivity}}
\centering
\renewcommand{\arraystretch}{0.75}
\setlength\tabcolsep{3pt}
\vspace{6pt}
\begin{tabular}{l c c c}
\hline
Mission & Payload launch mass & Mission cost & Reference \\
& (kg) & (\meuro) & \\
\hline
HST                & 11,110 & 8,037 & \cite{cost-hst-psp} \\
Chandra         &   5,860 & 4,114 & \cite{cost-chandra} \\
Cassini           &   5,820 & 2,806 & \cite{cost-planetary} \\
Cluster           &   4,800 &     944 & \cite{cost-cluster}  \\
Fermi             &   4,303 &     863 & \cite{cost-fermi} \\
INTEGRAL     &   4,000 &     419 & \cite{cost-integral} \\
Curiosity         &   3,893 & 2,590 & \cite{cost-planetary}  \\
XMM               &   3,800 & 1,113 & \cite{cost-xmm} \\
Juno               &   3,625 & 1,082 & \cite{cost-planetary} \\ 
Herschel         &   3,400 & 1,152 & \cite{cost-herschel} \\
RXTE              &   3,200 & 360 & \cite{cost-rxte} \\
SDO                &   3,100 & 865 & \cite{cost-sdo} \\
Rosetta           &   2,900 & 1,709 & \cite{cost-rosetta} \\
Galileo             &   2,560 & 1,275 & \cite{cost-planetary} \\
MAVEN            &   2,454 & 638 & \cite{cost-planetary}  \\
ROSAT            &   2,421 & 635 & \cite{cost-rosat}  \\
MRO                &   2,180 & 928 & \cite{cost-planetary} \\
GAIA                &  2,034 & 1,037 & \cite{cost-gaia} \\
Planck              &  1,900 & 775 & \cite{cost-planck}  \\
SoHO               & 1,850 &  1,469 & \cite{cost-soho}  \\
Suzaku             & 1,706 & & \\
AstroSat           & 1,515 &       27 & \cite{cost-astrosat}  \\
MMS                 & 1,360 &  1,054 & \cite{cost-gao}  \\
Venus Express & 1,270 &      300 & \cite{cost-vex}  \\
WIND               & 1,250 &  & \\
STEREO          & 1,238 &      614 & \cite{cost-stereo}  \\
Mars Express   & 1,223 &      374 & \cite{cost-mex}  \\
Dawn                & 1,218 &      439 & \cite{cost-planetary}  \\
Hipparcos         & 1,140 &      933 & \cite{cost-hipparcos}  \\
Kepler               & 1,052 &      636 & \cite{cost-kepler}  \\
GEOTAIL          & 1,009 &  & \\
Akari                 &     952 &      106 & \cite{cost-akari}  \\
Spitzer              &     950 & 1,188 & \cite{cost-spitzer}  \\
SWIFT              &     843 &      279 & \cite{cost-gallex-swift-timed}  \\
ACE                  &     752 &  & \\
InSight              &     694 &     714 & \cite{cost-planetary}  \\
PSP                  &     685 &  1,310 & \cite{cost-hst-psp}  \\ 
WISE                &     661 &     335 & \cite{cost-wise}  \\
TIMED              &    660 &      259 & \cite{cost-gallex-swift-timed}  \\
Double Star      &    560 & & \\
IMP-8               &    410 & & \\
NICER             &     372 &        53 & \cite{cost-nicer}  \\
NuSTAR          &     360 &      156 & \cite{cost-nustar} \\
TESS               &     325 &      275 & \cite{cost-tess}  \\
GALEX            &     280 &      120 & \cite{cost-gallex-swift-timed}  \\
DEMETER       &     130 &       21 & \cite{cost-demeter}  \\
\hline
\end{tabular}
\end{table}

\subsection{Ground-based observatories}

\begin{table}
\tiny
\caption{Construction and annual operating costs for ground-based astronomical observatories or 
telescopes that were considered in this work. If no cost estimates could be found the corresponding
entries in the table are blank.
\label{tab:groundactivity}}
\centering
\renewcommand{\arraystretch}{0.75}
\setlength\tabcolsep{3pt}
\vspace{6pt}
\begin{tabular}{l|c c c c}
\hline
Observatory & \multicolumn{4}{c}{Cost}  \\
& \multicolumn{2}{c}{Construction} & \multicolumn{2}{c}{Operations} \\
& (\meuro) & Reference & (\meuro\ yr$^{-1}$) & Reference \\
\hline
VLT (Paranal)                      & 1,384 & \cite{vanbelle2004} & 40 & \cite{opcost-vlt}  \\
ALMA                                  & 1,248 & \cite{cocost-alma} & 105 & \cite{opcost-alma}  \\
SOFIA                                 & 1,098 & \cite{cost-gao} & 90 & \cite{cost-gao}  \\
AAT                                     & 124 & \cite{cost-aat} & 15 & \cite{aat2009}  \\
VLA                                     & 345 & \cite{goodrich2019} & 10 & \cite{cost-vla}  \\
VLBA                                   & 132 & \cite{cost-vlba} & 15 & \cite{cost-vlba}  \\
IRAM                                   & 51 & \cite{cocost-iram} & 15 & \cite{opcost-iram}  \\
Gemini-South                      & 135 & \cite{goodrich2019} & 13 & \cite{goodrich2019}  \\
CFHT                                   & 85 & \cite{vanbelle2004} & 6.3 & \cite{cost-cfht} \\
ESO 3.6m (La Silla)            & 99 & \cite{cost-lasilla} & 5.2 & \cite{cost-lasilla}  \\
GBT                                    & 120 & \cite{cocost-gbt} & 10 & \cite{opcost-gbt}  \\
LOFAR                                & 200 & \cite{cocost-lofar} & 9.2 & \cite{opcost-lofar}  \\
JCMT                                  & 38 & \cite{cocost-jcmt} & 5.5 & \cite{opcost-jcmt}  \\
ATCA                                   & 95 & \cite{cocost-atca} & 3.5 & \cite{opcost-atca}  \\
H.E.S.S.                              & 49 & (1) & 8.8 & \cite{coutcomplet2016}  \\
MeerKAT                             & 128 & \cite{cost-meerkat} & 13 & \cite{cost-meerkat}  \\
GTC                                    & 125 & \cite{vanbelle2004} & &  \\
NRO                                    & 51 & (1) & 1.5 & (1)  \\
LMT                                     & 77 & \cite{leverington2016} & 3.1 & \cite{opcost-lmt}  \\
MLSO                                  & & & 1.2 & \cite{opcost-mlso}  \\
APEX                                  & 20 & (1) & 2.7 & \cite{olofsson2012}  \\
SMA                                    & 60 & \cite{masson1990} & &  \\
EHT                                    & 52 & \cite{cocost-eht} & - & -  \\
Noto Radio Observatory     & & & 1.5 & \cite{opcost-noto}  \\
2m TBL                               & 6.0 & (2) & 1.0 & \cite{opcost-tbl}  \\
2.16m (Xinglong Station)    & 7.3 & (2) & 0.7 & (3)  \\
1.93m OHP                         & 5.5 & (2) & 0.5 & (3)  \\
KMTNet                               & 17 & (1) & 1.7 & (1)  \\
THEMIS                               &       &       & 1.1 & (1) \\
2.4m LiJiang (YAO)             & 9.6 & (2) & 1.0 & (3)  \\
2m HCT (IAO)                     &  6.1 & (2) & 0.6 & (3) \\
1.5m Tillinghast (FLWO)      &  2.8 & (2) & 0.3 & (3)  \\
1.5m (OAN-SPM)                &  2.8 & (2) & 0.3 & (3)  \\
1.8m (BOAO)                      &  4.6 & (2) & 0.5 & (3) \\
1m (Pic-du-Midi)                  &  1.0 & (2) & 0.1 & (3) \\
1.3m Warsaw (OGLE)         &  2.0 & (2) & 0.2 & (3) \\
C2PU                                   &  2.0 & (2) & 0.2 & (3)  \\
TAROT                                 & 0.9 & (1) & 0.1 & (3)  \\
1m NOWT                            & 1.0 & (2) & 0.1 & (3)  \\
\hline
\end{tabular}
\newline
\footnotesize{
\begin{flushleft}
(1) private communication,
(2) derived using Eq.~(\ref{eq:cost}),
(3) assuming 10\% of construction cost as annual operation costs.
\end{flushleft}
}
\end{table}

Estimating the carbon footprint for ground-based observatories is more difficult than for space missions, 
since many ground-based observatories build up gradually, being complemented over the years with 
new telescopes or receiving upgrades for cameras, spectrographs, receivers, correlators and other 
instrumentation.
All these evolutions are difficult to reconstruct {\it a posteriori}, hence we take the approach to gather
as much information as possible, but without the requirement to be complete, leaving out information 
when it is not available, so that our estimate is formally only a lower limit to the true carbon footprint of 
ground-based observatories.
The results are summarised in Supplementary Table \ref{tab:groundactivity}.

In our list, there are a number of 1--2 metre class telescopes that are used by IRAP's stellar astrophysics 
group and for which reliable cost information is in general not available.
Several authors have however pointed out that construction cost scales with aperture size $D$, and for 
telescopes of modest size \cite{humphries1984} suggest that
\begin{equation}
\mbox{Construction cost} = 1.0 \times D^{2.58} \,\, \meuro\
\label{eq:cost}
\end{equation}
when converted to 2019 economic conditions, where $D$ is in units of metres.
Modern 1-metre class projects still satisfy this scaling law (e.g. \cite{liu2021}), hence we
adopt it for construction cost estimation for 1--2 metre class telescopes if no cost information
could be found.
In the absence of information on operating costs, we furthermore assume 10\% of
construction cost as annual operating costs, inferred from \cite{goodrich2019} for telescopes
with 1--10 \meuro\ construction cost.

We assembled construction costs and operation costs of observatories from public documents or via 
private inquiry.  
All cost values were corrected for inflation and converted to 2019 economic conditions.
For IRAM, we could not obtain construction costs for the initial installation, so we take the construction 
cost of the NOEMA upgrade as a lower limit to the total construction cost.  
We included the SOFIA airborne observatory in our list of ground-based observatories, since its lifecycle 
is closer to that of a ground-based observatory than a space mission.  
For the Event Horizon Telescope (EHT), which is a very long baseline interferometry array of existing 
millimeter and submillimeter wavelength facilities that span the globe \cite{akiyama2019}, we do not count 
the carbon footprint of the individual facilities, but the footprint related to the cost of combining the facilities 
into an Earth-wide VLBI project.  
For H.E.S.S., the operating costs are full costs extrapolated from the values provided by \cite{coutcomplet2016} 
for France and assuming a French contribution of 28\% to the project, estimated from the fraction of French 
authors on the paper \cite{hess2019}.

\section*{Contribution to the carbon footprint of IRAP}

To assess the carbon footprint of astronomical research infrastructures that can be attributed to 
an individual research institute we select our own institute IRAP as a case study.  
As a first method of attribution, we use the number of peer-reviewed scientific publications that have 
authors affiliated to IRAP. 
This is the method used to assess the scientific productivity of our institute in official evaluations.
For this estimate, we used the same ADS queries that we used for the determination of the number of 
papers for a given research infrastructure, but now we restricted the query to publications that included 
IRAP among the affiliations, to papers written in 2019, and to the facilities considered in this work.
Multiplying the results with the carbon intensity per paper for each space mission and ground-based
observatory and summing up the results gives a footprint of 20$\pm$4 \ktcoe\ (mass-based) to 
24$\pm$5 \ktcoe\ (cost-based) for the space missions and 13$\pm$5 \ktcoe\ for the ground-based 
observatories, totalling to 35$\pm$7 \ktcoe\ for IRAP.

We note, however, that research infrastructures that started operating only recently have a relatively 
large carbon intensity due to the short period over which the construction footprint is
distributed, leading to an artificially inflated carbon footprint attribution.  
We therefore use an alternative attribution method where we multiply the annual carbon footprint 
of an infrastructure with the fraction of peer-reviewed papers in 2019 that have authors affiliated
to IRAP.  
This results in a footprint of 12$\pm$2 \ktcoe\ (mass-based and cost-based) for the space missions and 
8$\pm$3 \ktcoe\ for the ground-based observatories, totalling to 20$\pm$3 \ktcoe\ for IRAP in 2019.
Attributing this footprint equally to the 144 astronomers with PhD degree that worked at IRAP in 2019 
results in 139$\pm$23 \tcoe\ per astronomer.  
If we instead divide the annual research infrastructure carbon footprint equally by the total number 
of staff that worked at IRAP in 2019 (263 people), we obtain 76$\pm$12 \tcoe\ per IRAP staff member.

We want to point out that this attribution method does not provide IRAP's share of the total carbon 
footprint of research infrastructures among all existing astronomical institutes in the world.  
Since scientific articles are often signed by authors from multiple institutes, each of these institutes 
will get the same attribution, implying that the sum of all attributions will exceed the total carbon
footprint of all research infrastructures.
The share can however be estimated by replacing the number of peer-reviewed papers by the
number of unique authors, i.e.~multiply the annual carbon footprint of an infrastructure with the 
fraction of unique authors of peer-reviewed papers in 2019 that are affiliated to IRAP.  
This results in a carbon footprint of 2.5$\pm$0.5 \ktcoe\ (mass-based) and 2.8$\pm$0.6 \ktcoe\ 
(cost-based) for the space missions, and 1.3$\pm$0.5 \ktcoe\ for the ground-based observatories, 
totalling to 4.0$\pm$0.7 \ktcoe\ for IRAP in 2019.  
For each of the IRAP astronomers with PhD degree this corresponds to a footprint of 27.4$\pm$4.8 \tcoe, for 
each person working at IRAP in 2019, the footprint is 15.0$\pm$2.6 \tcoe.  
We note that also in this share some double-counting may occur since an individual may be affiliated to 
multiple institutes. 
But for a given individual, the computation of the share should be accurate.

We point out that this result only covers the research infrastructures considered in this paper.
We assembled the list of infrastructures in our sample by scrutinising the titles and abstracts of the 2019
papers affiliated to IRAP, potentially missing infrastructures that were only mentioned in the body text of 
the papers.  
Our ADS queries, however, scan the full text of the papers, and hence would lead to a larger contribution 
to the IRAP footprint if more research infrastructures were covered.  
To test this hypothesis, we also estimate the IRAP 2019 contribution for all active space missions in
2019, based on the payload mass, and for all active optical-near-infrared (OIR) telescopes with at least 
3 metres of diameter.
The full list of facilities that we considered is provided in the Supplementary Data file that is available for 
download at \url{https://doi.org/10.5281/zenodo.5835840}.
In both cases, the extension of the list of infrastructures led to an increase of about 30\% in the attributed 
carbon footprint.  
If the same holds for other active infrastructures not included in this paper, the footprint of each IRAP 
astronomer with PhD degree would increase to about 36 \tcoe.

\section*{Is offsetting a solution?}

A possible solution that is often mentioned in the context of carbon footprint reductions is offsetting.
Briefly, offsetting is a process where someone else is paid to reduce his/her carbon footprint in place
of yours.  
Offsetting is already implemented in the European Union Emission Trading Scheme (EU ETS) but 
so far its impact on reducing the EU GHG emissions is at most modest \cite{bayer20, lehne21}.  
At the time of writing the paper (September 2021), the EU ETS carbon price was at 61 \euro\ 
per tonne of \co\ \cite{carbonprice}, which expressed in allowed \co\ emissions per amount of 
money spent corresponds to 16,400 \moneyeft.  
Individual offsetting projects have prices of the same order of magnitude, for example atmosfair, 
a popular scheme in Germany, uses 23 \euro\ per tonne of \co\ at the time of writing the
paper \cite{atmosfaire}, corresponding to an allowance of 43,500 \moneyeft.  
In other words, the allowance to emit \co\ is about two orders of magnitude larger than the 
actual emissions for a given amount of money (see the emission factors in Table 1).
Therefore the economic incentive to reduce \co\ emissions by offsetting is basically zero.

Is it possible to take a given amount of \co\ out of the atmosphere for 100 times less money 
than was spent for emitting it?  
As an example, we consider reforestation, which is a popular target of offsetting plans.  
Reforestation of one hectare of forest costs about 3 \keuro\ \cite{reforestation}.  
Depending on the prior use of the land, transforming agricultural land to forest brings an additional
carbon sequestration of 1.6 \tcoe\ per hectare and year, replacing a meadow by a forest brings 
an additional sequestration of 370 \kgcoe\ per hectare and year \cite{basecarbone}.  
In general, using agricultural land for reforestation should be avoided as the land is
needed for food production, creating in particular strong ethical issues when offsetting plans 
are implemented in the Global South.
Instead, we consider the sequestration potential of forests replacing meadows.  
Absorbing for example the ALMA operations footprint of 26 \ktcoeyr\ would require planting 
70,000 hectares of forest, which would correspond to an investment of 210 \meuro.  
Additionally absorbing its construction footprint of 300 \ktcoe\ over a decade (the
time scale suggested by the IPCC \cite{ipcc2021} to reach net zero GHG emissions) needs an 
additional 81,000 hectares of forest, corresponding to an investment of 240 \meuro, resulting 
in a total investment of 450 \meuro\ for offsetting ALMA's GHG emissions.  
This is about half of ALMA's construction costs, and requires reforestation of an area that is 
about the size of Monaco.  
The true price for offsetting is therefore significant, commensurable with the cost to build an 
astronomical research infrastructure, and it requires the availability of large areas of land 
that are not yet used otherwise.
Real offsetting (i.e.~the kind of offsetting we just described) might be a viable temporal solution 
if other GHG reduction schemes cannot be implemented, but we emphasise it requires 
considerable investment, which would need to factored into project costs as early as possible.

\end{addendum}

\end{document}